\documentclass[conference]{IEEEtran}
\IEEEoverridecommandlockouts
\usepackage{cite}
\usepackage{amsmath,amssymb,amsfonts}
\usepackage{algorithmic}
\usepackage{graphicx}
\usepackage{textcomp}
\usepackage{xcolor}
\usepackage{subfigure}
\usepackage[colorlinks,
            linkcolor=blue,
            anchorcolor=blue,
            citecolor=blue]{hyperref}
\usepackage{indentfirst}
\usepackage{mathrsfs}

\usepackage{multirow}
\usepackage[ruled]{algorithm2e}
\usepackage{multirow} 
\usepackage{bm}

\def\BibTeX{{\rm B\kern-.05em{\sc i\kerwn-.025em b}\kern-.08em
    T\kern-.1667em\lower.7ex\hbox{E}\kern-.125emX}}

\title{Rate-Splitting Multiple Access for Short-Packet Uplink Communications: A Finite Blocklength Analysis}

\begin{document}
%\thanks{J. Xu, O. Dizdar and B. Clerckx are with Imperial College London, London SW7 2AZ, UK (email: j.xu20, o.dizdar, b.clerckx@imperial.ac.uk). This work has been partially supported by the U.K. Engineering and Physical Sciences Research Council (EPSRC) under grant EP/N015312/1,EP/R511547/1.} 
%\author{Jiawei Xu, Onur Dizdar, \textit{Member, IEEE}, Bruno Clerckx, \textit{Fellow, IEEE}}
%\date{}
\author{

\IEEEauthorblockN{Jiawei~Xu, Onur~Dizdar, \IEEEmembership{Member, IEEE}, and~Bruno~Clerckx, \IEEEmembership{Fellow, IEEE}\vspace{-3.2em}}
\thanks{J. Xu is with Imperial College London. B. Clerckx is with the Department of Electrical and Electronic Engineering at Imperial College London, London SW7 2AZ, UK and with Silicon Austria Labs (SAL), Graz A-8010, Austria. O. Dizdar is with VIAVI Solutions UK Limited. (email: {j.xu20,b.clerckx}@imperial.ac.uk, onur.dizdar@viavisolutions.com).
%This work has been partially supported by the U.K. Engineering and Physical Sciences Research Council (EPSRC) under grant EP/N015312/1, EP/R511547/1.
}
}
\maketitle

\begin{abstract}
In this letter, we investigate the performance of Rate-Splitting Multiple Access (RSMA) for an uplink communication system with finite blocklegnth (FBL). Considering a two-user Single-Input Single-Output (SISO) Multiple Access Channel (MAC), we study the impact of blocklength and target rate on the throughput and error probability performance of RSMA where one user message is split. We demonstrate that RSMA can outperform Non-Orthogonal Multiple Access (NOMA) in terms of throughput and error probability performance.
\end{abstract}

\section{Introduction}

\par Rate-Splitting Multiple Access (RSMA) has been proven to be a strong multi-user transmission scheme and reliable interference management strategy for multi-antenna wireless networks\cite{clerckx2016rate,joudeh2016sum,mao2018rate}, and has the potential to tackle numerous challenges of modern communication systems\cite{mao2022rate}. RSMA relies on Rate-Splitting (RS) at the transmitters and Successive Interference Cancellation (SIC) at the receivers. It has been shown to bridge and outperform existing multiple access schemes, such as Space-Division Multiple Access
(SDMA), Non-Orthogonal Multiple Access (NOMA), Orthogonal Multiple Access (OMA) and multicasting under perfect and imperfect Channel State Information at the Transmitter (CSIT). In the downlink, the benefit of RSMA lies in its flexibility to partially decode the multi-user interference and partially treat it as noise; in contrast with SDMA, which fully treats the interference as noise, and NOMA, which fully decodes the interference\cite{clerckx2016rate,joudeh2016sum,clerckx2019rate,mao2018rate,hao2015rate}.
\par A majority of recent studies on RSMA consider downlink multi-antenna multiple-access scenarios. However, it has been demonstrated that uplink RSMA can achieve the optimal rate region of an $M$-user Gaussian Multiple Access Channel (MAC) using up to 2$M$-1 virtual point-to-point Gaussian channels created by message splitting\cite{rimoldi1996rate}. In \cite{zhu2017rate,yang2020sum,liu2020rate,zeng2019ensuring,tegos2022performance}, uplink RSMA was shown to lead to better outage probability, sum-throughput, sum-rate performance, high minimum date rate and lower latency compared to NOMA and OMA. %In \cite{zeng2019ensuring}, considering a max-min fairness problem in an uplink SISO setting, Single-Input Multiple Output (SIMO)-RSMA was shown to achieve higher minimum data rate and lower latency than SIMO-OMA and SIMO-NOMA.
\par A major application area for uplink multiple-access schemes is Ultra-Reliable Low-Latency  Communications (URLLC). The low-latency requirements in URLLC force systems to operate with short blocklength, bringing out the necessity to analyze the system performance with Finite Blocklength (FBL) codes. In the pioneering work \cite{polyanskiy2010channel}, the authors have laid the fundamental limits for the achievable rate with FBL codes for given blocklength and error probability, leading the path for theoretical performance analysis with FBL codes. % \cite{haghifam2017joint} examined the trade-off between sum rate and error probability and high throughput is achieved by optimizing the per-user error probability. 
Motivated by the trade-off between achievable rate and error probability, the error probabilities have been derived for NOMA in Additive White Gaussian Noise (AWGN) channel and Rayleigh fading channel with FBL in \cite{sun2018short,yu2017performance,dosit2019performance,schiessl2020noma}, and NOMA was shown to achieve higher effective throughput and reduced latency compared to OMA. %\cite{schiessl2020noma} studied the queueing delay of NOMA compared to OMA in the uplink with FBL and derived the closed-form approximations for the error probability and the result shows NOMA with jointly decoding is suitable for low-latency systems. 
The authors in \cite{dos2021rate} have shown RSMA can outperform OMA and NOMA in a network slicing scenario.
\par Nevertheless, all existing works on uplink RSMA are based on the Shannon theorem with infinite blocklength (IFBL) except [17] which applies RSMA to URLLC uplink transmission with network slicing but it focuses on the data rate with a fixed outage probability. To the best of our knowledge, this letter investigates sum-throughput and error probability the performance of uplink RSMA with FBL coding in a two-user system for the first time. We derive the analytical expressions of the error probability and come up with Successive Convex Approximations-based (SCA) algorithms to solve those problems. Our numerical results demonstrate that uplink RSMA achieves more reliable communication with FBL codes compared to uplink NOMA. 
%\par \textit{Organizations}: Section.\ref{2} introduces the system model of both RSMA and NOMA. Error probability expressions are given in Section.\ref{3}. In Section.\ref{4}, the problem of maximizing the sum-throughput is formulated and the algorithm is discussed. The results are shown in Section.\ref{5}. Finally, the paper is concluded in Section.\ref{8}.
%\par \textit{Notations}: $\mathcal{CN}(0,\sigma_n^2)$ represents a complex Gaussian distribution with mean $0$ and variance $\sigma_n^2$. $|\cdot|$ is the absolute value.

\section{System Model} \label{2}

\subsection{RSMA for Uplink Communications}\label{2.1}

\begin{figure}[t]
    \centering
    \includegraphics[width=3in]{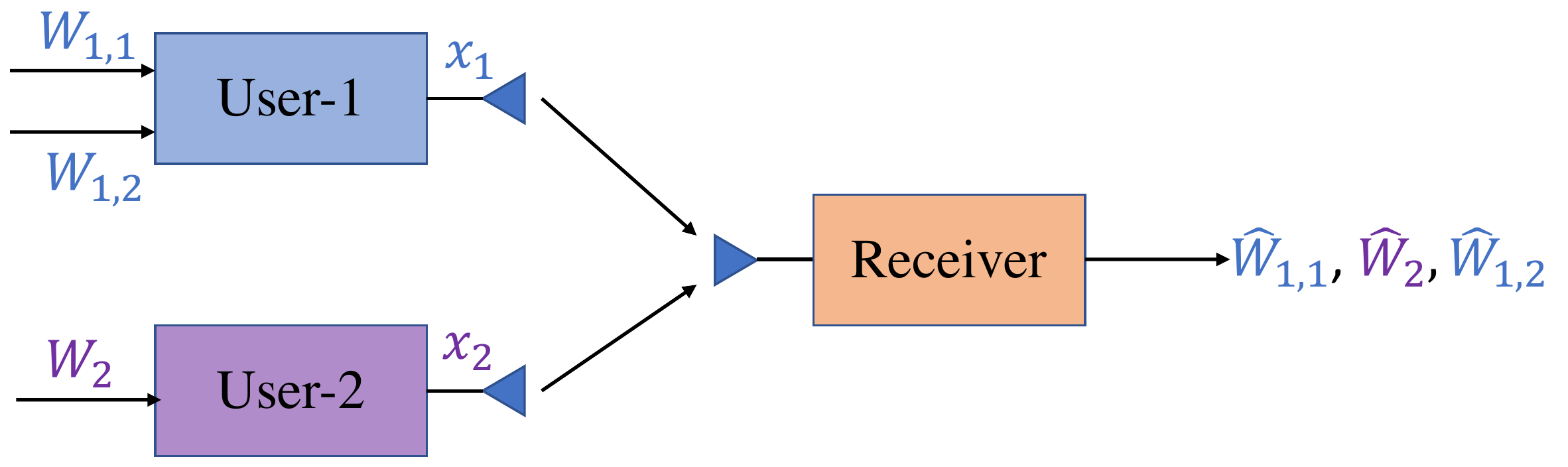}
    \caption{RSMA for two-user SISO MAC}
    \label{Fig.1}
\end{figure}

In this section, we consider a two-user uplink scenario with perfect CSIT and Channel State Information at Receiver (CSIR), where two single-antenna users indexed by $\mathcal{K}=\{1,2\}$ communicate with a single antenna Base Station (BS). Following uplink RSMA principle in \cite{rimoldi1996rate}, one user's message is split, resulting in a total of three streams transmitted to the BS. Fig.\ref{Fig.1} illustrates the two-user uplink RSMA example for the SISO MAC. User-$1$ splits its message $W_{1}$ into two parts, $W_{1,i},i\in(1,2)$, and $W_{2}$ denote the message of user-$2$. These three messages $W_{1,1}$, $W_{1,2}$, and $W_{2}$ are encoded into streams $s_{1,1}$, $s_{1,2}$ and $s_{2}$, respectively. Use $P_{k}$ to denote the transmit power of user-$k, k\in\{1,2\}$, so the received signal is expressed as
\begin{equation}
    y = h_1\sqrt{ P_{1,1}}s_{1,1}+h_1\sqrt{P_{1,2}}s_{1,2}+h_2\sqrt{P_{2}}s_2+n, \label{eq:1}
\end{equation}
where $P_{1,i}$ is the transmit power of $s_{1,i}, i\in(1,2)$, respectively and $\sum_{i=1}^2P_{1,i}\leq P_{1}$. $n \sim\mathcal{CN}(0,\sigma_n^2)$ is the AWGN at the BS. All possible decoding orders at the BS can be classified into three cases, i.e., (\romannumeral1) $s_{1,1}\to s_{2}\to s_{1,2}$ (i.e. the BS decodes $s_{1,1}$ first, followed by $s_2$, and finally $s_{1,2}$), (\romannumeral2) $s_{1,1}\to s_{1,2}\to s_{2}$ and (\romannumeral3) $s_{2}\to s_{1,1}\to s_{1,2}$. \footnote{By exchanging the orders of $s_{1,1}$ and $s_{1,2}$, there are three other decoding orders which are ignored in this letter without loss of generality.} We choose the decoding order (\romannumeral1) to write the Signal-to-Interference-plus-Noise (SINR) sequently. 
Accordingly, $s_{1,1}$ is decoded first at the BS by treating other signals as noise. The SINR $\gamma_{1,1}^{1}$ of the first decoded stream $s_{1,1}$ is 
\begin{equation}
    \gamma_{1,1}^{1} =\frac{ P_{1,1}|h_1|^2}{P_{1,2}|h_1|^2+P_{2}|h_2|^2+\sigma_n^2},
\end{equation}
where superscript $1$ represents the index of the user with split message. Assuming that $s_{1,1}$ is successfully decoded, it is reconstructed into $\widehat{W}_{1,1}$ and subtracted from the original received signal $y$ to obtain $y^{\prime}$. The SINR of the second decoded stream $s_2$ is 
\begin{equation}
    \gamma_2^{1} =\frac{P_{2}|h_2|^2}{P_{1,2}|h_1|^2+\sigma_n^2}.
\end{equation}
Assuming that $s_2$ is successfully decoded to obtain $\widehat{W}_{2}$, it is reconstructed and subtracted from the received signal $y^{\prime}$. Assuming $s_{1,1}$ and $s_{2}$ are correctly decoded, the SINR of the last decoded stream $s_{1,2}$ is 
\begin{equation}
    \gamma_{1,2}^{1} =\frac{P_{1,2}|h_1|^2}{\sigma_n^2}.
\end{equation}
Finally, if $s_{1,2}$ is successfully decoded, the estimated message $\widehat{W}_{1}$ for user-$1$ is obtained by combining $\widehat{W}_{1,1}$ and $\widehat{W}_{1,2}$.

\subsection{NOMA for Uplink Communications}

\par NOMA is a particular instance of the system model of Section \ref{2.1}, since $0\leq P_{1,1} \leq P_{1}$, we can regard NOMA as a subset of RSMA. The messages $W_{1}$ and $W_{2}$ from user-1 and user-2 are encoded into $s_1$ and $s_2$, respectively. The received signal at the BS is expressed as $y=h_1\sqrt{P_1}s_1+h_2\sqrt{P_2}s_2+n$.
%\begin{equation}
%    y=h_1\sqrt{P_1}s_1+h_2\sqrt{P_2}s_2+n.
%\end{equation}
We assume the decoding order 'NOMA-1' as $s_1\to s_{2}$ to demonstrate the SINR sequently. Accordingly, the SINR of streams under the assumption of successful decoding are written as $\gamma_1^{N1} = \frac{P_{1}|h_1|^2}{P_{2}|h_2|^2+\sigma_n^2}$ and $\gamma_2^{N1} = \frac{P_{2}|h_2|^2}{\sigma_n^2}$.
%\begin{subequations}
%\begin{align}
%    & \gamma_1^{N1} = \frac{P_{1}|h_1|^2}{P_{2}|h_2|^2+\sigma_n^2}\label{eq:6a}\\
%    & \gamma_2^{N1} = \frac{P_{2}|h_2|^2}{\sigma_n^2},
%    \label{eq:6b}
%\end{align}
%\end{subequations}
where the superscript 'N1' stands for 'NOMA-1'.
%\begin{remark}
    %If $P_{1,1}$ in RSMA scheme is set to $P_{1}$ or 0, RSMA boils down to NOMA with decoding order of $s_{1}\to s_{2}$ or $s_{2}\to s_{1}$, respectively. Since $0\leq P_{1,1} \leq P_{1}$, we can regard NOMA as a subset of RSMA.
%\end{remark}

\section{Error Probability Analysis} \label{3}

The finite blocklength achievable rate expression is given by \cite{polyanskiy2010channel} 
\begin{equation}
    r \approx 0.5\log_2(1+\gamma)-\sqrt{\frac{V}{N}}Q^{-1}(\epsilon)\label{eq:7}
\end{equation}
%with
%\begin{equation}
%    V = \log_2^2(e)\biggl(1-(1+\gamma)^{-2}\biggl),
%\end{equation}
where $V = \log_2^2(e)\biggl(1-(1+\gamma)^{-2}\biggl)$ is the channel dispersion, $\epsilon$ is the error probability, $\gamma$ is the SINR of the stream, $N$ is the blocklength, and $Q$ is the Q-function \footnote{Q-function is the tail distribution function of the standard normal distribution. Normally, Q-function is defined as: Q(x)=$\frac{1}{\sqrt{2\pi}} \int_x^{\infty}e^{-\frac{u^2}{2}}du.$}. Based on \eqref{eq:7}, we can write the error probability for a given transmission rate as
\begin{equation}
    \epsilon = Q \biggl( \frac{0.5\log_2(1+\gamma)-r}{\sqrt{V(\gamma)/N}}\biggl)=Q(\gamma, r) \label{eq:9}.
\end{equation}

\subsection{RSMA}

\par  Let us denote the target rates of user-$1$ and user-$2$ as $r_{1}$ and $r_{2}$, respectively. Given that user-$1$'s message is split, the target rates of two split streams are given by $r_{1,1}=\beta r_1$ and $r_{1,2}=(1-\beta)r_1$, where $0\leq\beta\leq 1$ is the rate allocation factor. We can write the error probability for each event listed above as $Q(\gamma_{1,1}^1, r_{1,1})$, $Q(\gamma_2^1, r_2)$ and $Q(\gamma_{1,2}^1, r_{1,2})$, respectively. For decoding order (\romannumeral1), the decoding error occurs due to three events listed as \footnote{Recall that for $\widehat{W}_{1}$ to be equal to the original message $W_{1}$, both $s_{1,1}$ and $s_{1,2}$ have to be correctly decoded.}: a) $s_{1,1}$ is incorrectly decoded; b) $s_{1,1}$ is correctly decoded but $s_2$ is incorrectly decoded; c) $s_{1,1}$ and $s_2$ are both correctly decoded but $s_{1,2}$ is incorrectly decoded.
%\begin{subequations}
%\begin{align}
%    \epsilon_a^1 &= Q(\frac{0.5\log_2(1+\gamma_{1,1}^1)-r_{1,1}}{\sqrt{V(\gamma_{1,1}^1)/N}}) \label{eq:10a}\\
%    \epsilon_b^1 &= 
%    Q(\frac{0.5\log_2(1+\gamma_2^1)-r_2}{\sqrt{V(\gamma_2^1)/N}}) \label{eq:10b}\\
%    \epsilon_c^1 &= Q(\frac{0.5\log_2(1+\gamma_{1,2}^1)-r_{1,2}}{\sqrt{V(\gamma_{1,2}^1)/N}})\label{eq:10c}.
%\end{align}
%\end{subequations}
Thus, the overall error probabilities for the message of user-$1$ and user-$2$ can be calculated as
\begin{subequations}
\begin{align}
    \epsilon_1^1 &=\epsilon_a^1+(1-\epsilon_a^1)\epsilon_b^1+(1-\epsilon_a^1)(1-\epsilon_b^1)\epsilon_c^1 \label{eq:11a}\\
    \epsilon_2^1 &=\epsilon_a^1+(1-\epsilon_a^1)\epsilon_b^1.\label{eq:11b}
\end{align}
\end{subequations}

\subsection{NOMA}

\par There are two events for incorrect decoding of $s_2$ : a) $s_1$ is incorrectly decoded; b) $s_1$ is correctly decoded but $s_2$ is incorrectly decoded. The probabilities of the specified events are calculated by $\epsilon_a^{N1} = Q(\gamma_1^{N1}, r_1)$ and $\epsilon_b^{N1} = Q(\gamma_2^{N1}, r_2)$, respectively.
%\begin{subequations}
%\begin{align}
%    \epsilon_a^{N1} & = Q(\frac{0.5\log_2(1+\gamma_1^{N1})-r_1}{\sqrt{V(\gamma_1^{N1})/N}})\label{eq:12a} \\ 
%    \epsilon_b^{N1} & = Q(\frac{0.5\log_2(1+\gamma_2^{N1})-r_2}{\sqrt{V(\gamma_2^{N1})/N}}).\label{eq:12b}
%\end{align}
%\end{subequations}
Accordingly, the overall error probabilities for the message of user-$1$ and user-$2$ are given by $\epsilon_1^{N1} =\epsilon_a^{N1}$ and $\epsilon_2^{N1}=\epsilon_a^{N1}+(1-\epsilon_a^{N1})\epsilon_b^{N1}$.
%\begin{subequations}
%\begin{align}
%    \epsilon_1^{N1} &=\epsilon_a^{N1} \label{eq:13a} \\
%    \epsilon_2^{N1} &=\epsilon_a^{N1}+(1-\epsilon_a^{N1})\epsilon_b^{N1}. \label{eq:13b}
%\end{align}
%\end{subequations}

\section{Problem Formulation and Algorithm} \label{4}

\par According to Shannon capacity definition, any rate inside the capacity region can be achieved with arbitrarily small error probability under the assumption of IFBL while rate inside the capacity region cannot be achieved with an arbitrarily small error probability in FBL regime is possible. Therefore we investigate the sum-throughput achieved by RSMA and NOMA to examine whether RSMA can achieve a lower error probability than NOMA. We define the throughput of each user as $T_{k}=r_{k}(1-\epsilon_{k})$ where $\epsilon_{k}$ is the overall error probability of user-$k$. For a given target rate pair $(r_{1}, r_{2})$, we formulate an optimization problem to maximize the sum-throughput as
\begin{subequations}\label{P14} 
\begin{align}
    \max_{\mathbf{\mathbf{P},\beta}} \quad & T_
    {tot} = T_1+T_2\\
    \mbox{s.t.}\quad
    & P_{1,1}+P_{1,2}\leq P_t \\
    & P_2\leq P_t, \\
    & 0\leq \beta \leq 1,
\end{align}
\end{subequations}
where $\mathbf{P}=[P_{k,1}, P_{k,2}, P_{j}]$ represents the transmit power of each stream.
%Recall that there are three decoding orders for RSMA, so we also need to find the optimal decoding orders. Thus, we first fix the decoding to obtain the power and rate allocation factors and then exhaustively search the decoding order.
We propose a SCA-based algorithm to optimize the transmit power. %\cite{mao2020max} has introduced a SCA-based algorithm for downlink RSMA where cooperative user relaying is enabled in the system with IFBL. The SCA approach discussed in this section differs from \cite{mao2020max} because it takes FBL into consideration for uplink RSMA.
\par We consider the scenario where the message of user-$k$ is split. $P_{k,1}$ and $P_{k,2}$ represent the transmit power of $s_{k,1}$ and $s_{k,2}$, respectively and $P_{j}$ represents the transmit power of user-$j$ ($j \neq k$), which does not perform message splitting. We demonstrate the proposed algorithm over decoding order (i), written as : $s_{k,1} \rightarrow s_{j} \rightarrow s_{k,2} $. The sum-throughout is defined as $T_{tot} =T_k+T_j=r_k+r_j-\text{TP}$.
%\begin{equation}
%     T_{tot} =T_k+T_j=r_k+r_j-\text{TP},
%\end{equation}
According to the error probability expressions, we can rewrite TP as
\begin{equation}
    %\begin{split}
    \text{TP}
     = (r_k+r_j)(\epsilon_a^k+\epsilon_b^k-\epsilon_a^k\epsilon_b^k)+r_k\epsilon_c^k(1-\epsilon_b^k)(1-\epsilon_a^k). \label{eq:26}
    %\end{split}
\end{equation}
Since $r_k+r_j$ is a constant, we can ignore it in the following analysis and minimize the value of TP. 
Thus, Problem (\ref{P14}) translates into 
\begin{subequations}\label{Problem25} 
\begin{align}
    \min_{\mathbf{\mathbf{P},\beta}} \quad & \text{TP}\\
    \mbox{s.t.}\quad
    & P_{k,1}+P_{k,2}\leq P_t, k\in\mathcal{K} \\
    & P_j\leq P_t, j\neq k\in\mathcal{K} \\
    & 0\leq \beta \leq 1.
\end{align}
\end{subequations}

Problem (\ref{Problem25}) is not convex due to the non-convex terms, $\epsilon_k^k$ and $\epsilon_j^k$ in \eqref{eq:26}. We introduce slack variables $t$, $\bm{\theta}=[\theta_a, \theta_b, \theta_c]$ and $\bm{\rho}=[\rho_{k,1}, \rho_{j}, \rho_{k,2}]$. With the aid of new variables, Problem (\ref{Problem25}) is transformed into Problem (\ref{P27}).
\begin{subequations}\label{P27}
\begin{align}
    \min_{\mathbf{P},\bm{\theta,\rho},\beta} \quad & t\\
    \mbox{s.t.}\quad
    & (r_k+r_j)(\theta_a^k+\theta_b^k(1-\theta_a^k))+ r_k\theta_c^k(1-\theta_b^k)(1-\theta_a^k)\nonumber\\
    &\leq t, k\in\mathcal{K},j\neq k\in\mathcal{K} \label{eq:27b}\\
    & \epsilon_i^k \geq \theta_i, i\in{(a, b, c)}, k\in\mathcal{K} \label{eq:27c} \\
    & \frac{P_{k,1}|h_k|^2}{P_{k,2}|h_k|^2+P_{j}|h_j|^2+\sigma_n^2}\geq \rho_{k,1} \label{eq:27d},k\in\mathcal{K},j\neq k\in\mathcal{K} \\
    & \frac{P_{j}|h_j|^2}{P_{k,2}|h_k|^2+\sigma_n^2}\geq \rho_{j} \label{eq:27e}, k\in\mathcal{K}, j\neq k\in\mathcal{K}\\
    & \frac{P_{k,2}|h_k|^2}{\sigma_n^2}\geq \rho_{k,2}, k\in\mathcal{K} \label{eq:27f} \\
    & P_{k,1}+P_{k,2}\leq P_t, k\in\mathcal{K} \label{eq:27g} \\
    & P_j \leq P_j, j\neq k\in\mathcal{K} \label{eq:27h}\\
    & 0\leq \beta\leq 1, \label{eq:27i}
\end{align}
\end{subequations}
Problem (\ref{P27}) is still not convex because $\theta_a\theta_b, \theta_a\theta_c, \theta_b\theta_c$ and $\theta_a\theta_b\theta_c$ in \eqref{eq:27b} are not convex. Therefore, they are approximated at the point ($\bm{\theta^{[n]}}$) at iteration $n$ by the first-order Taylor approximation, which can be written as
\begin{subequations}
\begin{align}
    \theta_a\theta_b & \geq \theta_a\theta_b^{[n]}+(\theta_b-\theta_b^{[n]})\theta_a^{[n]}\triangleq \Psi^{[n]}(\theta_a, \theta_b) \\
    \theta_a\theta_c & \geq \Psi^{[n]}(\theta_a, \theta_c) \\
    \theta_b\theta_c & \geq \Psi^{[n]}(\theta_b, \theta_c) \\
    \theta_a\theta_b\theta_c & \geq \theta_a\theta_b^{[n]}\theta_c^{[n]}+(\theta_b-\theta_b^{[n]})\theta_a^{[n]}\theta_c^{[n]}+(\theta_c-\theta_c^{[n]})\theta_a^{[n]}\theta_b^{[n]} \nonumber \\
    & \triangleq \Omega^{[N]}(\theta_a, \theta_b, \theta_c).
\end{align}
\end{subequations}

For constraint \eqref{eq:27c}, recall from \eqref{eq:9} 
that %$\epsilon=Q(s(\rho))$, where $s(\rho)=\frac{(0.5\log_2(1+\rho)-r)\sqrt{N}}{\sqrt{1-\frac{1}{(1+\rho)^2}}}$. Q-function 
$\epsilon_i^k$ is not convex. Thus, we also approximate it around the point ($\bm{\rho^{[n]}}$) by the first-order Taylor approximation which is given by
\begin{equation}
\begin{split}
    \epsilon_a^k & \geq Q(s(\rho_{k,1}^{[n]}))+(\rho_{k,1}-\rho_{k,1}^{[n]})\frac{dQ(s(\rho_{k,1}^{[n]}))}{d\rho_{k,1}^{[n]}}\\
    & \triangleq\Phi^{[n]}(\rho_{k,1}), k\in\mathcal{K}, j\neq k\in \mathcal{K},
\end{split}
\end{equation}
%where $\frac{dQ(s(\rho_{k,1}))}{d\rho_{k,1}}$ is calculated by 
%\begin{subequations}
%\begin{align}
%    & \frac{dQ(s(\rho_{k,1}))}{d\rho_{k,1}} = -\frac{1}{\sqrt{2\pi}}\frac{ds(\rho_{k,1})}{d\rho_{k,1}}e^{-\frac{s(\rho_{k,1})^2}{2}},k\in\mathcal{K} \\
%    & \frac{ds(\rho_{k,1})}{d\rho_{k,1}}  = \frac{\sqrt{N}(0.5\rho_{k,1}^2+\rho_{k,1}-0.5\text{ln}(1+\rho_{k,1})+\beta r_k)}{(\rho_{k,1}^2+2\rho_{k,1})\sqrt{\rho_{k,1}}},k\in\mathcal{K}.
%\end{align}
%\end{subequations}
If the argument of Q-function is larger than 0, Q-function is convex, otherwise it is concave. To keep $\epsilon_a^k\geq\Phi^{[n]}(\rho_{k,1})$, we define an additional constraint given as
\begin{equation}
    s(\rho_{k,1})\geq 0\Leftrightarrow 0.5\log_2(1+\rho_{k,1})-\beta r_k\geq 0, k\in\mathcal{K} \label{eq:31}.
\end{equation}
Therefore, constraint \eqref{eq:27c} is approximated around the point ($\mathbf{\rho^{[n]}}$) at iteration $n$ as 
\begin{equation}
    \begin{split}
        \Phi^{[n]}(\rho_{k,1}) & \geq\theta_a, k\in\mathcal{K} \\
        \Phi^{[n]}(\rho_{j}) & \geq\theta_b, j\neq k\in\mathcal{K} \\
        \Phi^{[n]}(\rho_{k,2}) & \geq\theta_c, k\in\mathcal{K} \\
        0.5\log_2(1+\rho_{k,1})-(1-\beta)r_k & \geq 0, k\in\mathcal{K} \\
        0.5\log_2(1+\rho_{j})-r_j & \geq 0, j\neq k\in\mathcal{K} \\
        0.5\log_2(1+\rho_{k,2})-(1-\beta)r_k & \geq 0,k\in\mathcal{K}. \label{eq:32}
    \end{split}
\end{equation}
\par Constraints \eqref{eq:27d} and \eqref{eq:27e} are 
%equivalently written into Difference-of Convex forms, which are given by 
%\begin{equation}
%    \begin{split}
%        & P_{k,2}|h_{k}|^2+P_{j}|h_{j}|^2+\sigma_n^2-\frac{P_{k,1}|h_{k}|^2}{\rho_{k,1}}, k\in\mathcal{K}  \\
%        & P_{k,2}|h_{k}|^2+\sigma_n^2-\frac{P_{j}|h_{j}|^2}{\rho_{j}}, j\neq k\in\mathcal{K} \label{eq:33}.
%    \end{split}
%\end{equation}
%and by rewriting the concave part with the first-order Taylor approximations, the constrains \eqref{eq:27d} and \eqref{eq:27e} are 
respectively approximated around the point ($\mathbf{P}^{[n]}, \bm{\rho}^{[n]}$) at iteration $n$ by 
\begin{equation}
    \begin{split}
        & P_{k,2}|h_{k}|^2 +P_{j}|h_{j}|^2+\sigma_n^2 \\ 
        & -\frac{P_{k,1}|h_{k}|^2}{\rho_{k,1}^{[n]}}+(\rho_{k,1}-\rho_{k,1}^{[n]})\frac{P_{k,1}^{[n]}|h_{k}|^2}{(\rho_{k,1}^{[n]})^2}\leq 0, k\in\mathcal{K}  \\
        & P_{k,2}|h_{k}|^2+\sigma_n^2 \\
        & -\frac{P_{j}|h_{j}|^2}{\rho_{j}^{[n]}}+(\rho_{j}-\rho_{j}^{[n]})\frac{P_{j}^{[n]}|h_{j}|^2}{(\rho_{j}^{[n]})^2} \leq 0, j\neq k\in\mathcal{K}. \label{eq:34}
    \end{split}
\end{equation}

Based on the above approximation methods, the original non-convex problem is transformed into a convex one and can be solved using the SCA method. SCA solves the problem by approximating a sequence of convex sub-problems. At iteration $n$, based on the optimal solution $(P^{n},\mathbb{\theta}^{[n]},\bm{\rho}^{[n]})$ obtained from the previous iteration $n-1$, we solve the following problem:
\begin{equation} \label{Problem35}
    \begin{split}
        \min_{\mathbf{P},\bm{\theta,\rho},\beta} \quad & t\\
        \mbox{s.t.}\quad
        & (r_k+r_j)(\theta_a+\theta_b-\Psi^{[n]}(\theta_a, \theta_b))\\
        & +r_k(\theta_c-\Psi^{[n]}(\theta_b, \theta_c)-\Psi^{[n]}(\theta_a, \theta_c)\\
        & +\Omega^{[N]}(\theta_a, \theta_b, \theta_c)) \leq t, k\in\mathcal{K},j\neq k\in\mathcal{K} \\
        & \eqref{eq:27f}, \eqref{eq:27g}, \eqref{eq:27h}, \eqref{eq:27i}, \eqref{eq:31}, \eqref{eq:32}, \eqref{eq:34}. 
    \end{split}
\end{equation}

The SCA-based power allocation algorithm is outlined in Algorithm 1 and the value of $\beta$ is optimized by exhaustive search. $\tau$ is the tolerance of convergence. Since the solution of Problem (\ref{Problem35}) at iteration $n-1$ is a feasible point of Problem (\ref{Problem35}) at iteration $n$ and the transmit power constraints \eqref{eq:27g} and \eqref{eq:27h}, the objective function $t$ is monotonically increasing which implies that the convergence of this proposed SCA-based algorithm is guaranteed. This SCA-based algorithm can be applied to K-user system with some modifications since the expression of TP of K-user system has a same form as (9). By introducing the slack variable, we can still use SCA algorithm in K-user scenario.
\begin{algorithm}
\caption{Proposed SCA-based algorithm}
\LinesNumbered
\SetKwInput{kwInit}{Initialize}
\kwInit{$n\gets0,t^{[n]}\gets0,\mathbf{P}^{[n]},\bm{\theta^{[n]},\rho}^{[n]}$;}
\Repeat{$|t^{[n]}-t^{[n-1]}|\leq \tau$}
 {$n\gets n+1$; \\
 Solve problem (\ref{Problem35}) using $\mathbf{P}^{[n-1]}, \bm{\rho}^{[n-1]}$ and find optimal $t^*$, $\mathbf{P^*},\bm{\theta^*,\rho^*}$; \\
 Update $t^{[n]}\gets t^*, \mathbf{P}^{[n]}\gets \mathbf{P}^*, \bm{\theta^{[n]}\gets\bm{\theta^*}}, \bm{\rho^{[n]}\gets\bm{\rho^*}}$;}
\end{algorithm}
The worst-case of computational complexity of the SCA and exhaustive-based algorithm is $\mathcal{O}\left(\delta^{-1}\log(\tau^{-1})[K]^{3.5}\right)$ where $\delta\in (0,1)$ is the increment between two adjacent candidates of $\beta$.

\section{Numerical Results}\label{5}

\par In this section, we perform simulations for AWGN channel. Without loss of generality, we assume that the noise variance is 1 and $P_t=10$dB. Results of four different blocklength which are $N=250, 500, 1500$ and $2500$ bits are compared. To evaluate the sum-throughput achieved by RSMA and NOMA according to different target rate pairs, we choose three kinds of symmetric target rate pairs which are noted as low, middle and high inside the capacity region with IFBL, respectively. To make target rate pairs symmetric, we choose these three kinds of target rate pairs to be on three circles with different radius. Thus, the selected target rate pairs are on the circles with radius of 0.8, 1.2 and 1.4 corresponding to low, middle and high, respectively. Then, we compare the rate region of NOMA and RSMA with different blocklengths with an error probability threshold. 'NOMA-1' and 'NOMA-2' refer to the NOMA with the decoding orders $s_1\to s_2$ and $s_2\to s_1$, respectively. 'RSMA-1' and 'RSMA-2' refer to the case where the message of user-1 and user-2 is split, respectively. 

\begin{figure}[t]
    \centering
    \subfigure[NOMA without time-sharing]{
    \centering
    \includegraphics[width=3.1in]{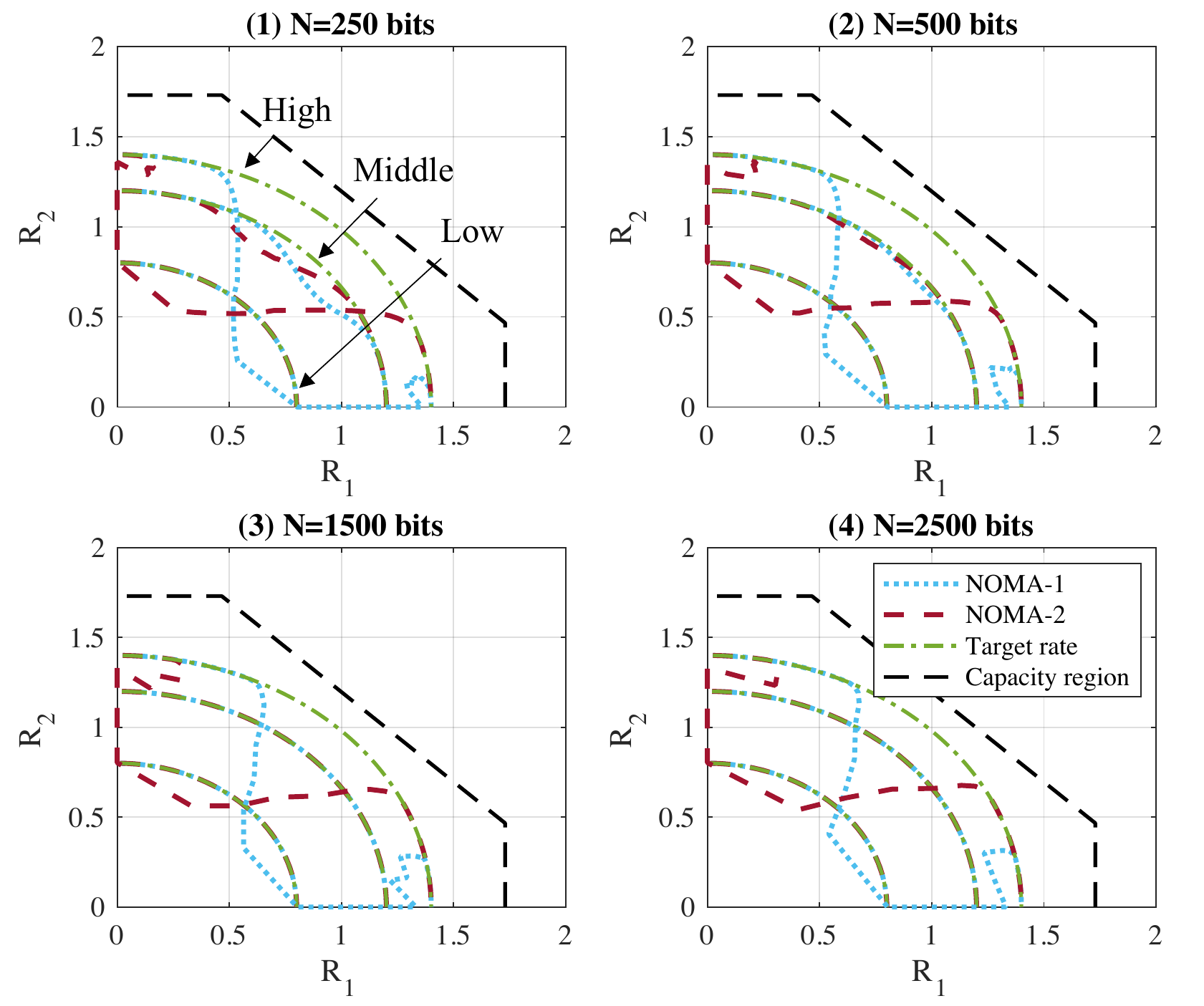}
    \label{Fig.3(a)}
    }\\
    \subfigure[RSMA without time-sharing]{
    \centering
    \includegraphics[width=3in]{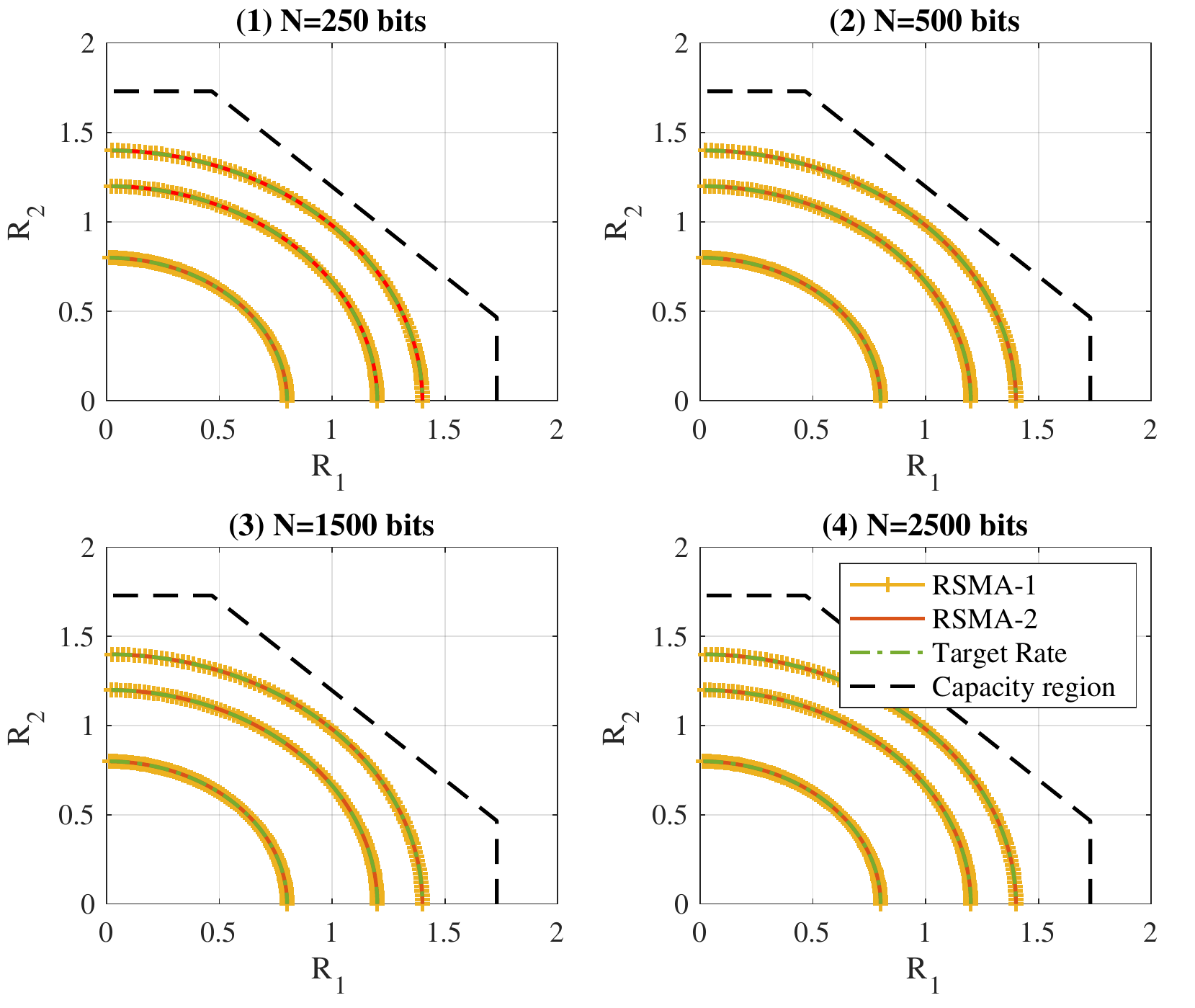}
    \label{Fig.3(b)}
    }
    \caption{Throughput of NOMA and RSMA without time-sharing}
    \label{Fig.3}
\end{figure}
\subsection{Error Probability}

\subsubsection{Without Time-Sharing}\label{5.1}
\par In Fig.\ref{Fig.3(a)}, NOMA can achieve the low and middle target rate pairs with 500 bits blocklength at least. For high rates (consider 'NOMA-1' only), when $r_1$ is small ’NOMA-1’ can achieve them with low error probability although $r_2$ is high since $P_2$ can be set high. As $r_1$ increases the performance deteriorates rapidly especially for user-2 even with large blocklength. We observe a little bump when $r_2$ approaches 0 in Fig.2(a) since when $r_2$ is less than a certain value, small $P_2$ can achieve $r_2$ and cause little interference to user-1. 
In Fig.\ref{Fig.3(b)}, we note that 'RSMA-1' and 'RSMA-2' have the same performance and RSMA can achieve the high target rate with very low error probabilities even with short blocklength without time-sharing.
%In Fig.\ref{Fig.4}, point M and N are two cross points of the rate region of 'NOMA-1' with blocklength equals to 500 and the high group of target rate pairs. For the target rate pairs which are at left of point M, 'NOMA-1' can achieve the target rate with low error probability because $r_1$ is small although $r_2$ is high and $P_2$ can be set very high to achieve $r_2$. As $r_1$ increase and $r_2$ decrease, we can increase $P_1$ and decrease $P_2$. Notice that transmitting at the target rate pairs outside the rate region of 'NOMA-1' would cause an obvious error probability.

%\begin{figure}
%    \centering
%    \includegraphics[width=3.5in]{NOMA_power_opt_N=500.png}
%    \caption{Throughput of user-1 and user-2 with N=500}
%    \label{Fig.4}
%\end{figure}

\subsubsection{With Time-Sharing}
\par
\begin{figure}
    \centering
    \includegraphics[width=3in]{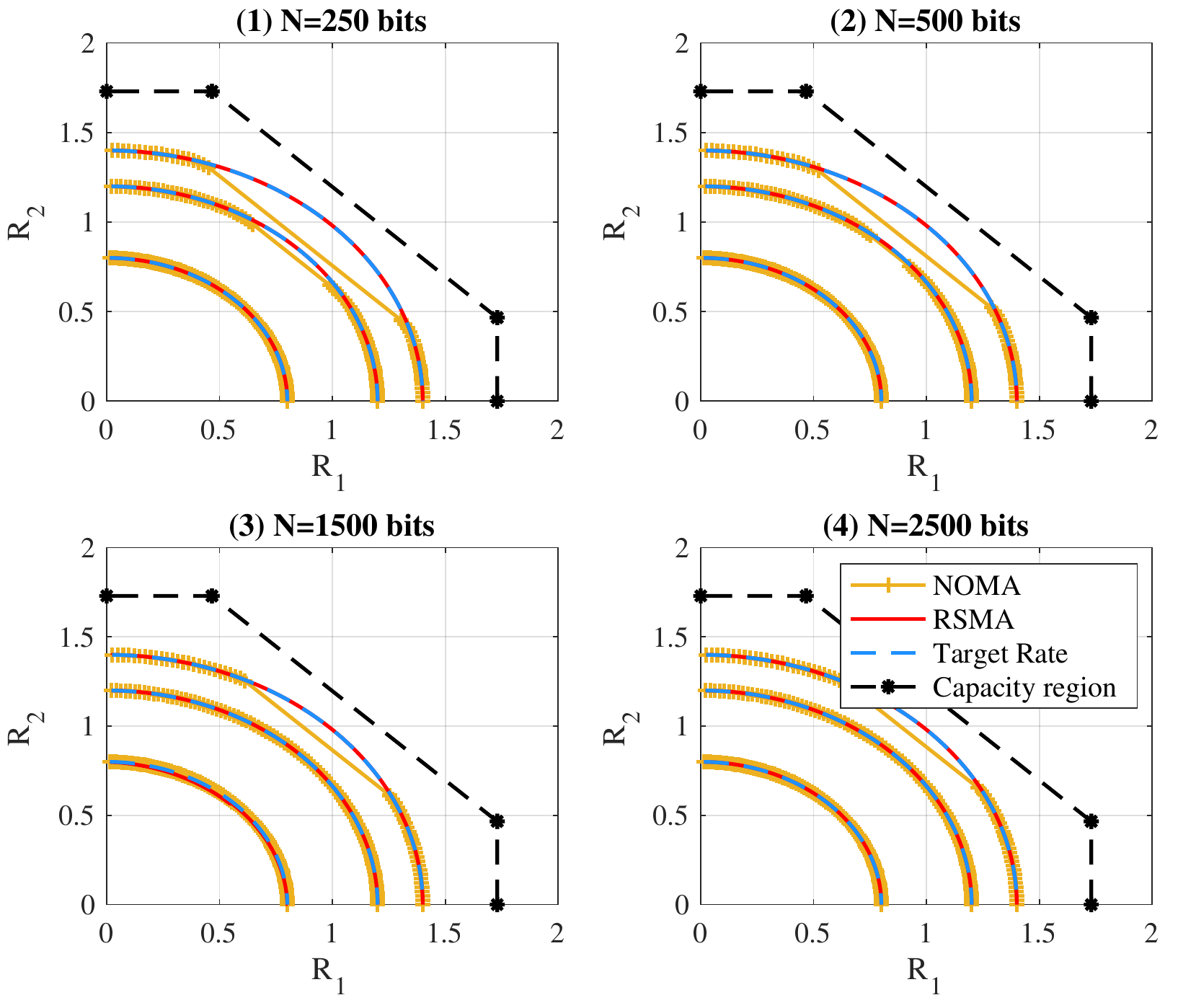}
    \caption{Throughput of NOMA and RSMA with time-sharing.}
    \label{Fig.5}
\end{figure}

  Compared to the result of no time-sharing in Fig.\ref{Fig.3(a)}, the performance of NOMA shown in Fig.\ref{Fig.5} becomes better and can achieve the same throughput as RSMA by transmitting at two target rate pairs with time-sharing. However RSMA still outperforms NOMA when target rate pairs are high. 

\subsection{Rate Region}\label{6}

\par Since 'NOMA-1 and 2' cannot achieve all the high target rate pairs in Sec.\ref{5.1},  we investigate the max target rate pairs that can be achieved by 'NOMA-1' and RSMA. In the simulation, we set the error probability $\epsilon=10^{-3}$ and we plot the rate region achieved by RSMA and 'NOMA-1' with this error probability constraint. 
\par
\begin{figure}[htbp]
    \centering
    \includegraphics[width=3in]{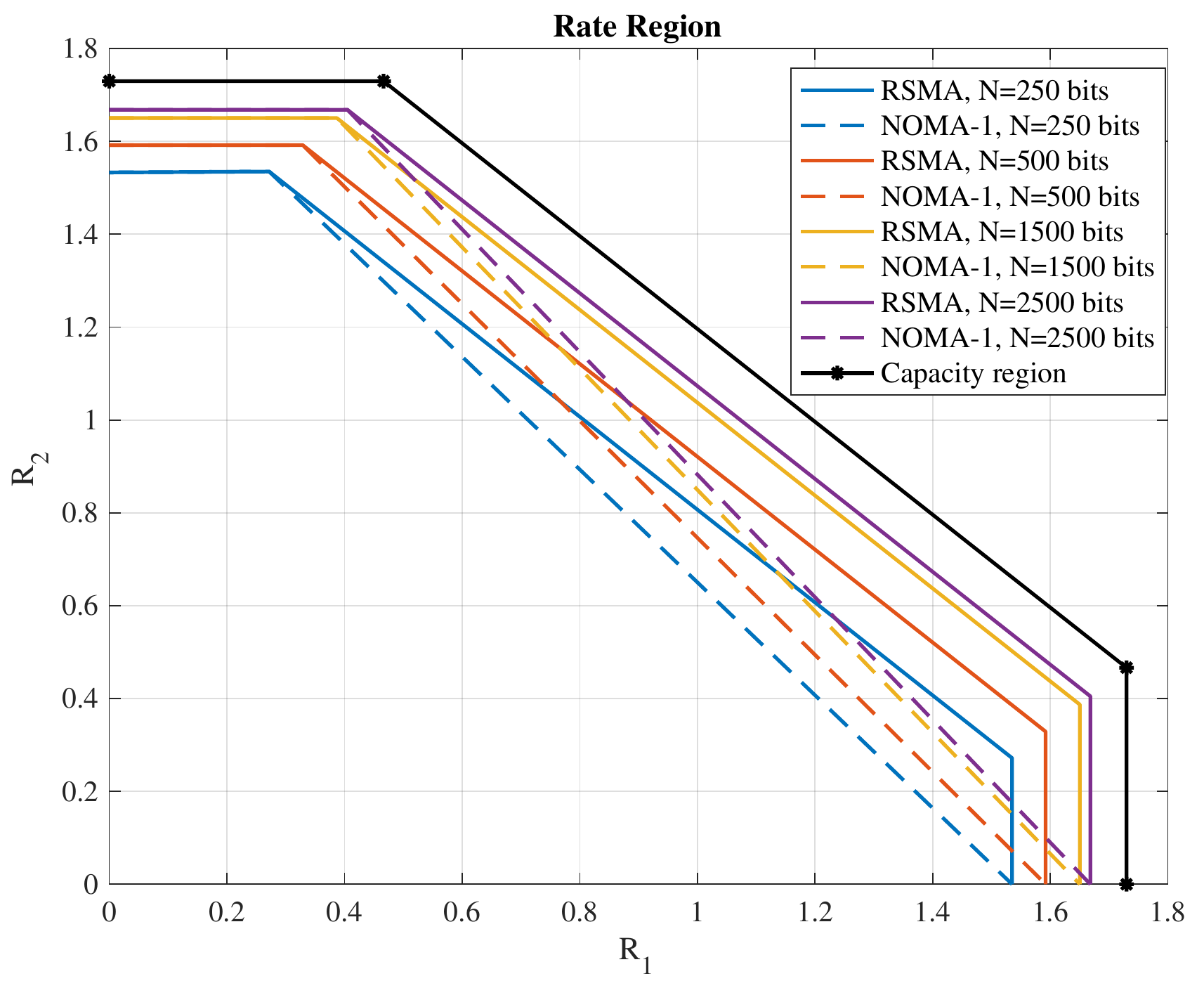}
    \caption{Rate region of user-1 and user-2 with $\epsilon=10^{-3}$. }
    \label{Fig.6}
\end{figure}

 In Fig.\ref{Fig.6}, one can observe that as the blocklength increases, the rate region of RSMA becomes closer to the capacity region while the rate region of 'NOMA-1' is always smaller than RSMA. It can be concluded that RSMA with decoding order of $s_{1,1}\to s_{2}\to s_{1,2}$ can achieve a larger rate region while NOMA can only achieve it by cooperating with time-sharing. The reason is that if we allocate all the transmit power to $s_{1,1}$, RSMA performs like 'NOMA-1' and if we allocate all the transmit power to $s_{1,2}$, RSMA performs like 'NOMA-2'. By changing the allocation of the transmit power between two split streams, RSMA bridges 'NOMA-1' and 'NOMA-2' without time-sharing.

\section{Conclusion}\label{8}
\par This letter studies whether RSMA can improve the error probability performance, sum-throughput and the rate region with FBL in SISO MAC. The numerical results show that RSMA can achieve significantly higher performance than NOMA without time-sharing. When time-sharing is included, NOMA can achieve the same sum-throughput as RSMA by transmitting at two different target rate pairs while RSMA can directly achieve it by transmitting at one target rate pair. When users transmit at the same target rate pair with RSMA and NOMA respectively, RSMA outperforms NOMA. Secondly, the max transmit target rate with constrained error probabilities is studied. The result shows that the rate region of RSMA is always larger than NOMA without time-sharing. In summary, when the target rates are fixed, RSMA provides lower error probabilities than NOMA and when the error probabilities are fixed, RSMA achieves a larger rate region than NOMA.

\bibliographystyle{ieeetr}

\bibliography{ref}

\begin{thebibliography}{10}

\bibitem{clerckx2016rate}
B.~Clerckx, H.~Joudeh, C.~Hao, M.~Dai, and B.~Rassouli, ``Rate splitting for
  mimo wireless networks: A promising phy-layer strategy for lte evolution,''
  {\em IEEE Communications Magazine}, vol.~54, no.~5, pp.~98--105, 2016.

\bibitem{joudeh2016sum}
H.~Joudeh and B.~Clerckx, ``Sum-rate maximization for linearly precoded
  downlink multiuser miso systems with partial csit: A rate-splitting
  approach,'' {\em IEEE Transactions on Communications}, vol.~64, no.~11,
  pp.~4847--4861, 2016.

\bibitem{mao2018rate}
Y.~Mao, B.~Clerckx, and V.~O. Li, ``Rate-splitting multiple access for downlink
  communication systems: bridging, generalizing, and outperforming sdma and
  noma,'' {\em EURASIP journal on wireless communications and networking},
  vol.~2018, no.~1, pp.~1--54, 2018.

\bibitem{mao2022rate}
Y.~Mao, O.~Dizdar, B.~Clerckx, R.~Schober, P.~Popovski, and H.~V. Poor,
  ``Rate-splitting multiple access: Fundamentals, survey, and future research
  trends,'' {\em arXiv preprint arXiv:2201.03192}, 2022.

\bibitem{clerckx2019rate}
B.~Clerckx, Y.~Mao, R.~Schober, and H.~V. Poor, ``Rate-splitting unifying sdma,
  oma, noma, and multicasting in miso broadcast channel: A simple two-user rate
  analysis,'' {\em IEEE Wireless Communications Letters}, vol.~9, no.~3,
  pp.~349--353, 2019.

\bibitem{hao2015rate}
C.~Hao, Y.~Wu, and B.~Clerckx, ``Rate analysis of two-receiver miso broadcast
  channel with finite rate feedback: A rate-splitting approach,'' {\em IEEE
  Transactions on Communications}, vol.~63, no.~9, pp.~3232--3246, 2015.

\bibitem{rimoldi1996rate}
B.~Rimoldi and R.~Urbanke, ``A rate-splitting approach to the gaussian
  multiple-access channel,'' {\em IEEE Transactions on Information Theory},
  vol.~42, no.~2, pp.~364--375, 1996.

\bibitem{zhu2017rate}
Y.~Zhu, X.~Wang, Z.~Zhang, X.~Chen, and Y.~Chen, ``A rate-splitting
  non-orthogonal multiple access scheme for uplink transmission,'' in {\em 2017
  9th International Conference on Wireless Communications and Signal Processing
  (WCSP)}, pp.~1--6, IEEE, 2017.

\bibitem{yang2020sum}
Z.~Yang, M.~Chen, W.~Saad, W.~Xu, and M.~Shikh-Bahaei, ``Sum-rate maximization
  of uplink rate splitting multiple access (rsma) communication,'' {\em IEEE
  Transactions on Mobile Computing}, 2020.

\bibitem{liu2020rate}
H.~Liu, T.~A. Tsiftsis, K.~J. Kim, K.~S. Kwak, and H.~V. Poor, ``Rate splitting
  for uplink noma with enhanced fairness and outage performance,'' {\em IEEE
  Transactions on Wireless Communications}, vol.~19, no.~7, pp.~4657--4670,
  2020.

\bibitem{zeng2019ensuring}
J.~Zeng, T.~Lv, W.~Ni, R.~P. Liu, N.~C. Beaulieu, and Y.~J. Guo, ``Ensuring
  max--min fairness of ul simo-noma: A rate splitting approach,'' {\em IEEE
  Transactions on Vehicular Technology}, vol.~68, no.~11, pp.~11080--11093,
  2019.

\bibitem{tegos2022performance}
S.~A. Tegos, P.~D. Diamantoulakis, and G.~K. Karagiannidis, ``On the
  performance of uplink rate-splitting multiple access,'' {\em IEEE
  Communications Letters}, vol.~26, no.~3, pp.~523--527, 2022.

\bibitem{polyanskiy2010channel}
Y.~Polyanskiy, H.~V. Poor, and S.~Verd{\'u}, ``Channel coding rate in the
  finite blocklength regime,'' {\em IEEE Transactions on Information Theory},
  vol.~56, no.~5, pp.~2307--2359, 2010.

\bibitem{sun2018short}
X.~Sun, S.~Yan, N.~Yang, Z.~Ding, C.~Shen, and Z.~Zhong, ``Short-packet
  downlink transmission with non-orthogonal multiple access,'' {\em IEEE
  Transactions on Wireless Communications}, vol.~17, no.~7, pp.~4550--4564,
  2018.

\bibitem{yu2017performance}
Y.~Yu, H.~Chen, Y.~Li, Z.~Ding, and B.~Vucetic, ``On the performance of
  non-orthogonal multiple access in short-packet communications,'' {\em IEEE
  Communications Letters}, vol.~22, no.~3, pp.~590--593, 2017.

\bibitem{dosit2019performance}
E.~Dosit, M.~Shehab, H.~Alves, and M.~Latva-aho, ``Performance of
  non-orthogonal multiple access under finite blocklength,'' {\em arXiv
  preprint arXiv:1902.09993}, 2019.

\bibitem{schiessl2020noma}
S.~Schiessl, M.~Skoglund, and J.~Gross, ``Noma in the uplink: Delay analysis
  with imperfect csi and finite-length coding,'' {\em IEEE Transactions on
  Wireless Communications}, vol.~19, no.~6, pp.~3879--3893, 2020.

\bibitem{dos2021rate}
E.~J. Dos~Santos, R.~D. Souza, and J.~L. Rebelatto, ``Rate-splitting multiple
  access for urllc uplink in physical layer network slicing with embb,'' {\em
  IEEE Access}, vol.~9, pp.~163178--163187, 2021.

\end{thebibliography}

\end{document}